\documentclass[english,russian,12pt,twoside]{article}
\usepackage{amssymb,amsmath}
\usepackage[russian]{babel}
\usepackage[cp866]{inputenc}
\begin{document}

\vspace{10mm}

\begin{center}

{\large\textbf {Newton's laws for a biquaternionic model of \\
the electro-gravimagnetic field,charges,currents\\ and their
interactions  }
 \vspace{4mm}

{ L.A.Alexeyeva}}

\vspace{4mm}

\textit{Wave Dynamics Laboratory, Institute of Mathematics,
Ministry of Education and Science\\ of the Republic
of Kazakhstan, 125 Pushkin Street, Almaty 050100, Kazakhstan,\\
E-mail:alexeeva@math.kz }
\end{center}
 \vspace{4mm}

\textbf{Abstract}

{\small{With use the Hamiltonian form of the Maxwell's equations
one biquaternionic model for  electro-gravimagnetic (EGM) field is
offered. The equations of the interaction of EGM-fields, which are
generated by different charge and current, are built. The field
analogs of three Newton's laws  are offered for free and
interacting charge-currents, as well as total field of
interaction. The invariance of these equations  at Lorentz
transformation  is investigated, and, in particular, of the
charge-current conservation law.
 It is shown that, by fields interaction, this law
differs from the well-known one. The new modification of the
Maxwell's equations is offered  with entering the scalar
resistance field in biquaternion of EGM-field tension.
Relativistic formulae of the transformation of density of the
masses and charge, current, forces and their powers are built. The
solution of the Cauchy problem is given for equation of
charge-current transformations.}} \vspace{10mm}

{\large\textbf{1 Introduction }} \vspace{4mm}

In the present paper, a biquaternionic model of the
electro-gravimagnetic (EGM) field is considered, which is called
\emph{energetic}. For this a  complex Hamiltonian form of Maxwell
equations (MEs) is used which allows to get the biquaternionic
form of these equations [1,2]. Note that Maxwell gave his
equations in quaternionic form, but the modern form belongs to
Heaviside [5]. Quaternionic forms of MEs were used by some authors
[5-7]  for their solving.  Kassandrov applied similar forms for
building a united field model [8].

Here we use scalar-vector form of biquaternion, which is very
impres\-sive  and  can be adap\-ted for writing the physical values
and equations. Based on   Newton's  laws, the biqua\-ter\-nio\-nic
trans\-formation equa\-tions of charges and currents at the presence
of the EGM fields are constructed. The relation of these equations
to hydro\-dynamics equations is considered. The  energy
conser\-vations laws at the presence of the fields inter\-action is
found.

The Lorentz invariance of  equations  of energetic field
interaction is studied, and also the invariance of charge-current
conservation law. It is shown that by the charges-current
interaction this law differs from the well-known one for the
closed electromagnetic (EM) field. The new modification of the MEs
is proposed for open field. One has to  introduce the scalar
\emph{field of resistance}. The relativistic transformation
 of mass  and electric charge and current densities,
 acting forces and their powers are constructed.

\vspace{4mm}

{\large{\bf 2 Hamiltonian form of Maxwell equations}} \vspace{4mm}

Let ${\bf M}=R^{1+3} = \{ (\tau ,x) = (ct,x): \, t\in
R^1,\,x=(x_1,x_2,x_3)\in R^3 \}$
 denote the  Minkowski  space. In   ${\bf M}$
 the symmetric form of Maxwell equations
for electromagnetic field can be written  as [1]
 $$
\partial _\tau  A + i\,rotA + J = 0,
\eqno{(2.1)}$$
$$\rho  = div\,A , \eqno{(2.2)}$$
where $A$ is complex vector of EM field tensions:
$$A = A^E  + i\,A^H  = \sqrt \varepsilon \, E + i\,\sqrt \mu \, H.
\eqno(2.3)$$
Vectors $E$ and $H$    are   the tensions of electric
and magnetic fields , $\varepsilon$ and $\mu $ are constants of
electric conductivity and magnetic permeability of the medium.
Complex charge density
 $\rho$ and current density $J$   are expressed through the densities of electric
 and magnetic charges and currents by
$$
\rho  = \rho ^E/\sqrt{\varepsilon}  - i\,
 \rho ^H /\sqrt{\mu}, \quad J = \sqrt {\mu }\, j^E  - i\sqrt \varepsilon \, j^H ,
\eqno(2.4)$$
$$
\rho ^E  = \varepsilon \,div\,E,\quad \rho ^H  =  - \mu \,div\,H.
\eqno(2.5)$$
 Energy density $W$  and the Pointing vector $P$ of
A-field read:
$$
W = 0,5\left( {\varepsilon \left\| E \right\|^2  + \mu \left\| H
\right\|^2 } \right) =  0,5(A,\bar A),\quad
 P = c^{ - 1} E \times H =
0,5i\,[A,\bar A], \eqno(2.6)$$
 where $\bar  A = \sqrt \varepsilon
E - i\sqrt \mu  H$,  $c = 1/\sqrt   {\varepsilon  \mu  }  $ is the
speed of light. Here and hereinafter
$$(a,b)=\sum_{i=1}^3 a_i b_i,  \quad [a,b] =a \times b=\sum_{i,j,k=1}^3{
e_{ijk}e_i a_j b_k} $$
 are  scalar and vector products of the  $a$ and $b$
respectively,
 $e_{ijk}$ is Levi-Chivita pseudotensor and  $e_i\,\,(i=1,2,3)$ are the unit vectors
 of the
 cartesian coordinate system in $R^3$ .

One can see,  that energy density  $W$ is simply the module of the
complex vector \emph{A} (half of it). Note that,  differently from
MEs , all relations for A-field don't contain the constants of EM
medium. In particulary,  the velocity of electromagnetic waves in
this coordinate system is non-dimensional and  equal to 1.

Here some known statements  are given which are due to Maxwell
equations [1]. \vspace{2mm}

 \textbf{Theorem  2.1.} \textit{For given current and charge the solution of Eq.}
 (2.1)  \textit{satisfies to the wave equation}:
  $$     \Box \,A  =  (\partial  _\tau  ^2  - \Delta )A = i\,  rot\,J  -
 grad\rho  - \partial _\tau  J,
   \eqno(2.7)$$
  \textit{and  the conservation laws  of  charge and energy is hold}:
 $$     \partial _\tau  \rho  + div\,J = 0,
 \eqno(2.8)$$
 $$     \partial _\tau  W + div\,P =  - {\mathop{\rm Re}\nolimits} (J,\bar
                          A) = c^{-1}(j^H H - j^E E).
 \eqno(2.9)$$

In MEs the density of the magnetic charges  $\rho ^H  = 0$.This
means that the  magnetic field is solenoidal one: $div\,H=0$.

But it is known that the classical gravitation is scalar field. It
can be described by a scalar gravitational potential, which
depends on the masses. Here  these two fields are united in a
unique  \emph{gravimagnetic field}. It is possible to do so if to
introduce a density of gravitational mass in Maxwell equations. In
particular the following hypothesis can be proposed.\vspace{3mm}

\textbf{H y p o t h e s i s. }The \emph{density $\rho ^H $ is
equal to density of gravitational mass}.\vspace{3mm}

Hereinafter we will show that this hypothesis has a theoretical
acknowledgements which bring us very  plausible effect.

Thence follows that potential part of $H$ describes the gravitational
field and solenoidal part describes magnetic field. So $H$-field is \emph{gravimagnetic
field}.
 Consequently, $A$-field is \emph{electro-gravimagnetic}.
 Since its dimensionality is defined by density of energy, it possible be called as \emph{energetic field}.

We name $j^H $ \textit{gravimagnetic} current. If $\rho ^H
= 0$  it is \textit{magnetic} current. If $H$-field is potential ($rot H=0$), $j^H $ is
\textit{mass} current.

Note, the system of MEs is unclosed. It allows for given
 charge and current to define the A-field, or  for given A-field
  to find corresponding charges and currents. If they are unknown
   then for its closing usually  the equations of mechanics of mediums are used.
    However we will enter here the other image, using
    biquaternionic form of these equations and Newton laws.

For going  to this form and new equations we will give the
thumbnail sketch on functional space of biquaternions and
operation on it. \vspace{6mm}

{\large{{ \textbf{ 3  Differential algebra of biquaternions:
Bigradients}}}} \vspace{4mm}

 The functional space of biqua\-ter\-nions is the
space of complex quaternions:
  $${\bf K} ({R}^{1+3}) = \{
{\bf  F} = f(\tau,x ) + F(\tau,x ) ,\}$$
 where  $f$ is complex
function and  $F$ is three-dimensi\-onal vector-func\-tion with
complex compo\-nents. We assume further that $f$ and $F$ are local
inte\-grable  and differen\-tiable on \textbf{M}.

The space \textbf{K} is \emph{associative }but non commutative
algebra with addition
$${\bf  F}  +  {\bf  G} =  (f
+  g)  + (F  + G),$$ and product( $ \circ$)
$$
{\bf F} \circ {\bf G} = (f + F) \circ (g + G) = (fg - (F,G)) +
                 (fG + gF + [F,G]).
\eqno(3.1)$$
The biquaternion ${\bf \bar F}  =  \bar f +\bar F$
is called \textit{complex conjugated},  ${\bf F}^{*}  =  \bar f -
\bar F$ is \textit{conjugated}.   If  ${\bf   F}^*    = {\bf F}$,
it is called \textit{selfconjugated}. The example of
selfconjugated biquaternion is $\textrm{{\bf F}}=f +iF$, when f
and F are real functions.\vspace{2mm}

\textbf{Definition 3.1. }\emph{Scalar product} of $ {\bf F}_1
,{\bf F}_2 $ is  defined by
$$ \left( {{\bf F}_1 ,{\bf F}_2 } \right) = f_1 f_2
+ \left( {F_1 ,F_2 } \right).  $$

\textbf{Definition 3.2. }\emph{Norma} of {\bf F} is defined by
 $$ \left\|
{\bf F} \right\| = \sqrt {\left( {{\bf F},{\bf \bar F}} \right)}
= \sqrt {f \cdot \bar f + \left( {F,\bar F} \right)}  = \sqrt
{\left| f \right|^2  + \left\| F \right\|^2 }. $$

\textbf{Definition 3.3.} \emph{Pseudonorma} of {\bf F} is defined
by
$$
\left\langle {\bf F} \right\rangle  = \sqrt {f \cdot \bar f -
\left( {F,\bar F} \right)}  = \sqrt {\left| f \right|^2  - \left\|
F \right\|^2 }.
$$

Hereinafter the \emph{mutual complex gradients} are used:
$$\nabla^ +   = \partial _\tau   + i\nabla
,\quad \nabla^ -
                   = \partial _\tau   - i\nabla,$$
where $\nabla  = grad = (\partial _1 ,\partial _2 ,\partial  _3  )$.
 The action of these differential operators on \textbf{K} is determined as in the biquaternions algebra :
(accordingly to a sign)
 $$
   \nabla^ \pm  {\bf F} = (\partial _\tau   \pm i\nabla ) \circ (f +
   F) = (\partial _\tau  f \mp i\,(\nabla ,F)) \pm\partial _\tau  F \pm
                      i\nabla f \pm i[\nabla ,F]=
  $$
  $$ = (\partial _\tau  f \mp i\,div\,F) \pm \partial _\tau  F \pm igrad\,f
                            \pm i\,rot\,F.$$
Further we call them \emph{bigradients.} \vspace{6mm}

{\large{{\textbf{4 Biwave equations: Cauchy problem}}}}
\vspace{4mm}

It is easy to check that wave operator can be  presented in the
form
$$ \Box=\frac{\partial^2}{\partial\tau^2}-\triangle= \nabla^ -   \circ \nabla^ +
= \nabla^ +   \circ \nabla^ -  .$$
 Using this property,  it is possible to build the solution of the differential equations  of the type:
$$ \nabla^ \pm  {\bf K} = {\bf G}.
\eqno(4.1)$$
We call such equations the \emph{biwave equations}.
 From (4.1) it is follow that
 $$\Box {\bf K} = \nabla^ \mp {\bf G}.$$
 Its solution is  the
convolution
$${\bf K} = \nabla^ \mp  {\bf G} * \psi, \eqno(4.2)$$ where   $\psi
(\tau ,x)$ is the fundamental solution of the wave equation
$$\Box\psi  = \delta (\tau) \delta(x).$$
 This solution is also the solution of
 (4.1). Really, using property of differentiation of convolution, we have
\[
\nabla^ \pm  {\bf K} = \nabla^ \pm  \nabla^ \mp  \left( {{\bf
G} * } \right.\left. \psi  \right) = \Box\left( {{\bf G} * }
\right.\left. \psi  \right) = \left( {{\bf G} * } \right.\left.
\Box{\psi } \right) = {\bf G*}\delta (\tau )\delta (x) = {\bf G}.
\]

The fundamental solutions are defined for constructing solutions
of  wave equation. For value problems it is convenient to use as
fundamental solution the simple
 layer on light cone$\tau  = \left\| x \right\|$:
$$\psi  = (4\pi \left\| x \right\|)^{ - 1}
\delta (\tau  - \left\| x \right\|).$$
 In this case, as it is easy to show, by writing the convolution
in integral type, the decision (4.2) will be equal to zero
for $\tau=0$. We use this for building of the solution of (4.1)
with Cauchy data.\vspace{4mm}

\textbf{\emph{Cauchy problem.}} The initial data are given: ${\bf
K}(0,x)={\bf K_0}(x).$ It is require to find the solution of
(4.1), which satisfies these data. \vspace{2mm}

To solve the problem  we use here methods of distribution theory
[9]. We will consider regular generalized functions of the type
 ${\bf \widehat{G}}=H(\tau){\bf G}(\tau,x)$, where $H(\tau)$  is the Heaviside function.
 By using differentiation of generalized function we obtain
  $\nabla^ \pm  {\bf \widehat{K}} = {\bf \widehat{G}}+\delta(\tau) {\bf {K_0}}(x)$.
Hence
$$
{\bf H(\tau)K}(\tau,x) = \nabla^ \mp  \{H(\tau){\bf \widehat{G}} *
\psi\}+{\bf G}(0,x)
 \mathop *\limits_x \psi+\nabla^ \mp\{{\bf K_0}(x) \mathop *\limits_x \psi\}
 \eqno(4.3)$$
where the sign "$\mathop *\limits_x$ " means that convolution is
given only over $x$.
 Its integral form is
$$
{4\pi }{\textbf{K}}(\tau,x) =- \nabla^ \mp  \left\{\int\limits_{r
\le \tau } {\frac{{{\bf G}(\tau  - r,y)}}{r}} dV(y)
+\tau^{-1}\int\limits_{r =\tau } {\bf K_0}(y)
dS(y)\right\}-\tau^{-1} \int\limits_{r =\tau } {\bf G}(0,y) dS(y),
 \eqno(4.4)$$
where $ r=\|y-x\|,\,dV(y)=dy_1dy_2dy_3$,  $dS(y)$ is a
differential of sphere's area.

This formula is a generalization of the famous Kirchhoff formula  for solution of
 Cauchy problem  for wave equation [9].
 \vspace{4mm}

{\large{{\textbf{ 5 Biquaternions of A-field}}}} \vspace{4mm}

We introduce the following biquaternions:
$$\textit{potential}\quad{\bf\Phi } = i\phi - \Psi ,$$
$$ \textit{tension}\quad \textbf{A}=0+A,$$
$$\textit{ charge-current density  }\quad
{\bf \Theta } = -i\rho - J,$$
 $$\textit{energy-pulse density }\quad{\bf \Xi }= 0,5\,{\bf A}^* \circ {\bf A}  = W + iP .$$
In the biquaternions space the Maxwell  equations (2.1)-(2.2) have the
simple form [2]
$$
\nabla^+  {\bf A} = {\bf \Theta } . \eqno(5.1)$$ If the potential
satisfies  to  Lorentz calibration
$\partial_\tau \phi  - div\,\Psi  = 0$,
 then
$${\bf A} = \nabla^ -  {\bf \Phi } .$$
If we  take corresponding complex gradient, we get the wave
equations
$$
\Box{\bf \Phi } =   {\bf\Theta }, \eqno(5.2)$$
$$
 \Box {\bf A} =  \nabla^- {\bf \Theta }.
\eqno(5.3)$$
Hence it follows that the   bigradient from
A-field poten\-tial defines biqua\-ter\-nions, correspon\-ding to the
field tension, charge and current. The scalar part of
bigradient of energy-pulse gives the law of the energy conser\-vation
[2].

One can  see that the  \emph{charges and currents is  simply
physical appearance of the bigradient of EGM field }.
\vspace{2mm}

\emph{\textbf{Cauchy problem for Maxwell equations}}. From
equation (4.3) it follows that  for given charge-current and
initial data ${\bf A}(0,x)={\bf A_0}(x)$,
 the solution of (5.1) is given by
\[
{4\pi }{\textbf{A}} =- \nabla^ -  \left\{\int\limits_{r \le \tau
} {\frac{{{\bf \Theta}(\tau  - r,y)}}{r}} dV(y) +
\tau^{-1}\int\limits_{r =\tau } {\bf A_0}(y) dS(y)\right\} -
\tau^{-1} \int\limits_{r =\tau } {\bf \Theta}(0,y) dS(y).
 \]
Hence it is easy to write the  integral representations for vector
of the EGM-field tension  \emph{E,H}. \vspace{10mm}

{\large{{\textbf{6   Lorentz transformation on M}}}}
\vspace{4mm}

Denote
 $$ {\bf Z} = \tau  + ix,\quad {\bf \bar Z} = \tau  - ix .$$
 It is easy to see that
\[
{\rm   }{\bf Z} = {\bf Z}^* ,\,\,{\bf \bar Z} = {\bf \bar Z}^*
,\,\,\left\| {\bf Z} \right\|^2  = \left\| {{\bf \bar Z}}
\right\|^2 = ({\bf Z},{\bf \bar Z}),\,\,\left\langle {\bf Z}
\right\rangle ^2  = \left\langle {{\bf \bar Z}} \right\rangle ^2 =
{\bf Z} \circ {\bf \bar Z}.
\]
Consider the selfconjugated biquaternions by using the hyperbolic
sine and cosine:
 $$ {\bf U} = \textrm{cosh}\,\theta  +
ie\,\textrm{sinh}\,\theta ,\,\,{\bf \bar U}  =
\textrm{cosh}\,\theta - ie\,\textrm{sinh}\,\theta ,\quad \left\| e
\right\| = 1.$$ Here $\theta$ is real number. It is easy to check
that
$$
{\bf U}\circ{\bf \bar U}=1 . \eqno(6.1)$$

   \textbf{ Lemma  6.1. } \emph{Classical Lorentz transformation } $
L:{\bf {\rm Z}} \to {\bf {\rm Z}}^{\bf '} $ \emph{has the form}
$$ {\bf Z}' = {\bf U} \circ {\bf Z} \circ {\bf U},\quad {\bf Z} =
{\bf \bar U}
  \circ {\bf Z}' \circ {\bf \bar U},$$

\textbf{Proof.} The direct calculation  proves  this lemma. The
pseudonorma is saved:  \[ \left\langle {{\bf Z'}} \right\rangle
^{\bf 2}  = {\bf U} \circ {\bf Z} \circ {\bf U} \circ {\bf \bar U}
\circ {\bf \bar Z} \circ {\bf \bar U} = \left\langle {\bf Z}
\right\rangle ^{\bf 2}.\] Here the property of associativity and
(6.1) we are used.

If we use the notations
 $$ ch2\theta  = \frac{1}{{\sqrt {1 - v^2 }
}},\quad sh2\theta  = \frac{v}{{\sqrt {1 - v^2 } }},\quad\left| v
\right| < 1 ,$$
 then the scalar and vector part of biquaternions can be written in the form of known relativistic formulas:
\[
\tau ' = \frac{{\tau  + v(e,x)}}{{\sqrt {1 - v^2 } }},\quad x' =
(x - e(e,x)) + e\frac{{(e,x) + v\tau }}{{\sqrt {1 - v^2 } }},
\]
\[
\tau  = \frac{{\tau ' - v(e,x)}}{{\sqrt {1 - v^2 } }},\quad x =
(x' - e(e,x')) + e\frac{{(e,x') - v\tau '}}{{\sqrt {1 - v^2 } }},
\]
It corresponds to the motion of coordinate system X in the direction
of vector \emph{е }with velocity \emph{v}.

\textbf{ Lemma 6.2.}  \emph{The conjugated  quaternions} $$ {\bf
W} = \cos \varphi  + e\sin \varphi ,\,\, {\bf W}^{\bf *}  = \cos
\varphi  - e\sin \varphi ,\,\,\, ( \left\| e \right\| = 1 )$$
\emph{define the group of transformation on}
\textbf{M} \emph{which are orthogonal on  vector part Z} :
$$ {\bf Z}^{\bf '}  = {\bf W} \circ {\bf Z} \circ {\bf W}^{\bf
*} ,\,\,\, {\bf Z} = {\bf W}^{\bf *}  \circ {\bf Z}^{\bf '}  \circ
{\bf W} .$$

It is rotation around the vector  \emph{e}  on the angle $2\varphi$ .
    As the result of these two lemmas we have the following.\vspace{2mm}

\textbf{Lemma 6.3.} \emph{The Lorentz transformation  on
\textbf{\textrm{\textbf{M}}} can be defined by }:
$${\bf Z}^{\bf '}  = {\bf L} \circ {\bf Z} \circ {\bf L}^{\ast} ,\,\,
{\bf Z} = \bar{{\bf L}^{\ast} } \circ {\bf Z}^{\bf '}  \circ \bar{{\bf L}},
\eqno(6.2)$$
$${\bf L} = {\bf W} \circ {\bf U} = ch(\theta  +i\varphi
) + iesh(\theta  +i\varphi ) ,\,\,\, {\bf L}^*  = {\bf U}^*  \circ
{\bf W}^*  = ch(\theta  - i\varphi ) + iesh(\theta  -i\varphi ),$$
\emph{The pseudonorm is saved for Lorentz transformation :
}$\langle\textbf{Z}\rangle=\langle\textbf{Z}'\rangle$.

It is easy to see that ${\bf \bar L} \circ {\bf L}^{*}= {\bf
L}^{*}\circ {\bf \bar L} = 1 $, because the pseudonorm
$\textbf{Z}$ is saved.\vspace{6mm}

{\large{{\textbf{7  Lorentz transformation of biwave equations}}}}
\vspace{4mm}

 Let us consider how bigradients are transformed under
Lorentz transformation.

\textbf{Lemma   7.1.} \emph{If} $ {\bf Z}^{\bf '}  = {\bf L} \circ
{\bf Z} \circ {\bf L}^* $, \emph{then} $$ \textrm{\textbf{D}}' =
{\bf \bar L}^*  \circ \nabla \circ {\bf L},\quad
\textrm{\textbf{D}} = {\bf L} \circ \nabla' \circ {\bf\bar L}^*
,$$ \emph{where} $ \textrm{\textbf{D}} = \nabla^ + $ \emph{or}  $
\textrm{\textbf{D}} = \nabla^ - .$

 Based on  this lemma, consider how the biwave equation (4.1) is changed by
Lorentz trans\-formation. Using associati\-vity of the product and
charac\-teristic of \textbf{L}, we get
\[
\nabla'{\bf K}' = \left( {{\bf \bar L}^*  \circ \nabla \circ
{\bf L}} \right)\left( {{\bf \bar L}^*  \circ {\bf K} \circ {\bf
L}} \right) = {\bf \bar L}^*  \circ \nabla \circ {\bf K} \circ
{\bf L} = {\bf \bar L}^* \circ {\bf G} \circ {\bf L} = {\bf G'}.
\]
Hence, the form of equation is saved:
$$
\left( {\frac{\partial }{{\partial \tau '}} \pm i\nabla '}
\right){\bf K'} = {\bf G'},
$$
where     $ {\bf K}^{\bf '}  = {\bf \bar L}^*  \circ {\bf K} \circ
{\bf L} ,\,\,\, {\bf G}^{\bf '}  = {\bf \bar L}^*  \circ {\bf G}
\circ {\bf L}. $ From here we have the following result. \vspace{2mm}

  \textbf{  Theorem  7.1.} \emph{The Lorentz transformation of the Maxwell
equations can be written as follows:}
$$\textbf{D}^+{\bf A'} = {\bf \Theta'},\,\,
\emph{where}    \,\, {\bf A}^{\bf '}  = {\bf \bar L}^*  \circ {\bf
A} \circ {\bf L} ,\,\,\, {\bf \Theta}^{\bf '}  = {\bf \bar L}^*
\circ {\bf \Theta} \circ {\bf L}. $$

\emph{Relativistic formulas for tension, charge and current (when
$\varphi=0$)}:
$$
A' =(A-e(e,A)) + e{\frac{{(e,A)  }}{{\sqrt {1 - v^2 } }}} \eqno(7.1)$$
$$
\rho' = \frac{{\rho - v(e,J)}}{{\sqrt {1 - v^2 } }},\quad  J' =
(J-e(e,J)) + e{\frac{{(e,J) - v\rho }}{{\sqrt {1 - v^2 } }}}
\eqno(7.2)$$

One can see that  the tension of A-field  always increases in
direction of  vector \emph{е}. In the absence of current,  the
charge-mass will increase. At the presence current, depending on
directions of their motion, the charge-mass can  increase or
decrease. \vspace{6mm}

{\large{{ \textbf{ 8.  The third Newton law: The power and density
of acting forces}}}}\vspace{4mm}

 Let us consider two EGM fields ${\bf A}$ and ${\bf
A}'$. Their generating charges and currents are ${\bf \Theta  }$ и ${\bf \Theta'}$.
 We call a \emph{power-force density  }  biquaternion
$$
{\bf F} = M - iF =   \;{\bf \Theta } \circ {\bf A}' =  -
                (i\rho  + J) \circ A' = (A',J) - i\rho A' + [A',J]
\eqno(8.1)$$
   which is acting from side of $A'$ -field on the charge and current of $A$-field.
    Really,  using (2.3) and (2.4), the scalar part is determined as
    power density of acting forces:
$$
 M = (A',J) = c^{ - 1} ((E',j^E ) + (H',j^H )) + i((B',j^E ) -
 (D',j^H ))
\eqno(8.2)$$ Selecting the real and imaginary parts of vector
form of biquaternion, we get expressions for density of acting
forces $\left( {F = F^H + i\,F^E } \right)$:
$$
 F^H= \rho ^E E' + \rho ^H H' + j^E\times B' - j^H  \times D'
\eqno(8.3)$$
$$
 F^E = c\left( {\rho ^E B' - \rho ^H D'} \right) + c^{ - 1}
\left( {E' \times j^E  + H' \times j^H } \right) \eqno(8.4)$$
Here  $B  =  \mu H$ is an analog of a vector to magnetic induction
(in torsional part complies with it),
 $ D = \varepsilon E$ is a vector of the electric offset.

The potentional part of $H$ describes the tension of gravitational
field. Torsional part of this vector describes magnetic field. The
scalar part of ${\bf \Theta },\,{\bf \Theta '}$ contains the
densities of electric charge and mass, its vector part contains
the densities of electric and mass currents.

Coming from these suggestions, in formula (8.3)  the known
forces are standing, consecu\-tively:  Coulomb force $\rho ^E E'$,
gravita\-tional force $\rho ^H H'$ (more exactly, it complies with
it in poten\-tial part \textit{H'}), Lorentz force  $j^E\times B'$
(more exactly,it complies with it in torsional part \textit{B'}))
and new force $-D' \times j^H $, which we call
\textit{gravielectric}.  In real part of the power (8.2) we
see the powers of Coulomb force, gravita\-tional and magnetic force.

The power of gravielectric force does not enter in real part of
(8.2)  as it does not work
 on the mass displacement, because it is perpendicular to its velocity. It is interesting that
  Lorentz force also does not enter in real part of (8.2). It proves that this force
   is perpendicular to mass velocity, though
directly from Maxwell equations this does not follow.

Naturally, in analogy, to expect that equations (8.4) describe forces, causing change of
electric current, but in imaginary part \textit{M} the power stands which corresponds to it. On the
virtue  of the third Newton law  about acting and counteracting forces, we suppose that must be
executed:
${\bf F'} =  - {\bf F}.$ From here we get    \vspace{4mm}

\textbf{\textit{  The law of fields action and reaction   }}:
$$
{\bf \Theta } \circ {\bf A'} =  - {\bf \Theta '} \circ {\bf A}.
\eqno(8.5)$$
 It is interesting to note that in scalar part it requires the equality of
the powers corresponding to forces, acting on charges and currents
of the other field. That is befitted  with what is known in
mechanics as the identity  reciprocity of Betti, which is usually
written for the work of forces.
 \vspace{4mm}

{\large{{\textbf{  9. The second Newton law: Transformations
equation}}}}\vspace{4mm}

 The  charge-current field is changed under influence
of the field of other charge and current. As it is well known,
direction of the most intensive change of the scalar field
describes its gradient. In analogy we can expect that the most
intensive change of the charge-current field  occurs toward its
bigradient. Naturally expect, that this change must occur
toward power-force, acting on the side of the second field on the
first one.  So the law of the change of the charge-current  field
under the action of the others (like second Newton law) is offered
in the manner of the following equations.

\textit{The equations of the charge-current interaction :}
$$
\kappa \nabla^ -  {\bf \Theta } = {\bf F} \equiv   {\bf \Theta
 } \circ {\bf A}',\,\,\,\,
\kappa \nabla^ -  {\bf \Theta }' =  {\bf \Theta }' \circ {\bf A},
\eqno(9.1)$$
$$
 {\bf \Theta } \circ {\bf A}' =  - {\bf \Theta }' \circ {\bf A},
\eqno(9.2)$$
$$
\nabla^ +  {\bf A} ={\bf \Theta } ,
                              \,\,\,\,
\nabla^ +  {\bf A}' = {\bf \Theta }' . \eqno(9.3)$$ Here
(9.1) correspond to the second Newton law which is written
for each charge-current of interacting field.  Equation (9.2) is
the third Newton law.  Together with Maxwell equations for these
fields (9.3) they give closed system of the nonlinear
differen\-tial equations for deter\-mination ${\bf A,
A',\Theta,\Theta'}$.     Entering the constant of  inter\-action
$\kappa $ is connected with dimensio\-nality. Revealing scalar and
vector part in (9.1), we have  \vspace{3mm}

\textbf{\textit{The equations of  charge-current transformations}}:
$$
 i\,\kappa \left( {\partial _\tau  \rho  +div\,J} \right) = M,
\eqno(9.4)$$
$$
i\,\kappa \left( {\partial _\tau  J -i\,rot\,J + \nabla \rho }
  \right) = F.
\eqno(9.5)$$ At first let consider  the second equation.    By
virtue (2.2), (2.3), (2.4), we obtain  analog of   \vspace{3mm}

\textbf{\textit{The second Newton law for charge-current field:} }
$$
\kappa  \left(\sqrt \varepsilon\,\partial _\tau  j^H  +\sqrt \mu
 \, rot\,  j^E+\mu  ^{  - 0,5} grad\,\rho  ^H
\right)=\rho ^E E' + \rho ^H H' + j^E  \times B' - j^H \times D',
\eqno(9.6)$$
$$
 \kappa  \left(   \sqrt
\mu\,\partial _\tau  j^E  -\sqrt \varepsilon  \, rot\,j^H
+\varepsilon ^{ - 0,5}  grad\,\rho  ^E \right)= c\left( {\rho ^E
B' - \rho ^H D'} \right) + c^{ - 1} \left( {E' \times j^E  + H'
\times j^H } \right). \eqno(9.7)$$
In (9.6) the value
$\kappa \sqrt  \varepsilon   j^H  $ is an analog of linear
momentum. Equation (9.7) describes the influence of the
external field on electric charge.

If one field is much stronger  the second one, then it is possible
to neglect the second field change under influence of charge and
current from first field. In this case we get closed system of
equations for determination  the charge and current motion of the
first field under action of second field:$$\kappa \nabla^ - {\bf
\Theta } ={\bf \Theta } \circ {\bf A}',$$ where  ${\bf A}'$ is
given. The corresponding A-field is defined by Maxwell equations (9.3).
 \vspace{6mm}

{\large{{\textbf{ 10. First Newton law: Free field}}
}}\vspace{4mm}

 Let us consider A-field, which is generated by
$\bf{\Theta}$, in the absence of other charges-currents. We call
it a \emph{free field}. In this case ${\bf F}=0$. From
(9.1) we get \emph{inertia law},
 which is analog of \vspace{3mm}

 \textbf{\textit{The first  Newton law  for charge-current field:}}
$$
  \nabla^ -  {\bf \Theta } = {\bf 0},
\eqno(10.1)$$
which is equivalent to equations:$${\partial _\tau
\rho  +div\,J}  = 0,\,\,\,\,
      \partial _\tau  J -i\,\, rot\,J+\nabla \rho   = 0.
$$
For initial designations we have following formulas:
$$
  \partial _t \rho ^E  + div\,j^E  = 0,\quad \partial _\tau  j^E
          = \sqrt {\varepsilon /\mu }\, rot\,j^H  - c\,grad\,\rho ^E,
\eqno(10.2)$$
$$
\partial _t \rho ^H  + div\,j^H  = 0,\quad \partial _\tau  j^H
        =   -\sqrt {\mu /\varepsilon }\, rot\,j^E  -c\,grad\,\rho ^H.
\eqno(10.3)$$ Consequently, charge-current conservation law
(2.8) holds in the absence of the external field.\vspace{3mm}

\textbf{\emph{Cauchy problem}.} For free field  the solution of
this problem  is given by  the formula:
$$
\kappa{\bf \Theta}(\tau,x) = \kappa\nabla^ {-}\{{\bf
{\Theta}}_0(x) \mathop *\limits_x \psi\}= - \frac{\kappa
H(\tau)}{4\pi } \nabla^ {-}\left\{ \tau^{-1}\int\limits_{r =\tau
} {\bf \Theta}_0(y) dS(y)\right\},
 \eqno(10.4)$$
and tensions of  A-field are defined in Section 5.  \vspace{6mm}

{\large{{\textbf{11 Modified Maxwell equations: Scalar resistance
field}}}}\vspace{3mm}

 Let us consider the first equation (9.4).
Evidently, it is the charge-current conservation law, which
contains   the power of external acting forces \emph{M} in the
write-hand site. When \emph{M}=0, this law has well-known form
(2.8), which we have had for Maxwell equation (see Theorem 2.1).

This  means that by EGM fields  interaction we must enter the
scalar part  in tensions biqua\-ter\-nion:
 $${\bf A}=i\,a(\tau,x)+A(\tau,x).$$
We call $a(\tau,x)$ the \emph{A-field resistance}. From
(9.1) -(9.3) follows that
$$\Box{\bf A} =\nabla^ -{\bf \Theta } =\kappa^{-1}{\bf F}.$$
The scalar part of this is
 $$ \,\kappa  \Box \,a =iM.$$

\textbf{Remark 11.1. }In system of Maxwell equations
(2.1),(2.2) the first
 equation defines the currents, the second one  determinate the
 charge, but  the charges conservation law is due to these two
 equations. It can be get, if we  take divergence  in (2.1) with provision for (2.2).
  However, the biquaternionic approach, as it is  shown here ,
  brings to modification of the Maxwell equations, which,
   in what follows from (9.3), has a following type:\vspace{3mm}

 \textbf{  \emph{The modified Maxwell equations }}
$$
J=grad\, a -\partial _\tau  A - i\,rotA,\quad \rho  = div\,A
-\partial _\tau a. \eqno(11.1)$$

If $\rho$ and $J$ are  known, this system for
 determining $a$ and $A$ is closed. Only in closed system
 (in the absence of external field) $a=0$ and it has the
 classical type  (2.1)-(2.2).

Obviously, by introducing the \textit{resistance scalar field}
$a$, the form of scalar and vector
 parts of power-force biquaternion (8.1) is changed, as follows
$$
{\bf F} =  {\bf \Theta } \circ {\bf A}' = ((A',J)+a'\rho)-i(a'J
+\rho A') + [A',J]. \eqno(11.2)$$
We can see the   additional
summands which appears in the presence of powers ($a'\rho$) and
force ($-ia'J$). The vector $a'J$ is called a \emph{ resistances
force} of $A'$-field.

Selecting real and imaginary part
 of this vector    we  get additional summands in expressions
   for density of  electric $F^E$ and gravimagnetic $F^H$ forces, forming
    $F\,\quad( F = F^H + i\,F^E )$ in  (9.5) with provision
     for the resistance force of fields, which we must  add in
      right parts of the Eqs. (9.6), (9.7).\vspace{3mm}

\textbf{\emph{Cauchy problem  for equation of transformation.
}}Using formula (4.2), we get
$$
\kappa{\bf \Theta}(\tau,x) = \nabla^ {+} \{ H(\tau){\bf
{F}}(\tau,x) * \psi\}+{\bf {F }}(0,x) \mathop *\limits_x
\psi+\kappa\nabla^ {+}\{{\bf {\Theta}}(0,x) \mathop *\limits_x
\psi\}
 \eqno(11.3)$$
This  equation gives  the system of integral equations for
determining ${\bf \Theta}$, as the right-hand side contains   ${\bf
\Theta}$ in  \textbf{F}. It can be used for solving the problem if
 we neglect the second field change. In general case, we write the
similar equation  for second field ${\bf \Theta'}.$ These two
equations give us  the full system of integral equations for
determination of charges and currents by their interaction, if
initial  state of fields are known. \vspace{3mm}

\textbf{\emph{{ The Lorentz transformations of  Transformation
Equations. }}}  According to Theorem  7.2,
 Lorentz transformations for $\textbf{A}, \,{\bf \Theta },\, \textbf{F}
$  have such form:
$$
 {\bf A}^{\bf '}  = {\bf \bar L}^*  \circ {\bf A} \circ {\bf
L} ,\,\,\, {{\bf \Theta }}^{\bf '}  = {\bf \bar L}^*  \circ \bf
\Theta  \circ {\bf L},\,\,\, {{\bf {F}}}^{\bf '}  = {\bf \bar L}^*
\circ {{\bf \textsc{F}}} \circ {\bf L}. \eqno(11.4)$$
(here the  sign '     means the coordinates in moving coordi\-nate system).
  Note, Lorentz transfor\-mation of power-force density at the presence of
  inter\-action of two fields of form  (8.1) have
the same form:
$$
{\bf F'} = {\bf \Theta'_1 } \circ {\bf A'_2} ={\bf \bar L}^* \circ
{\bf \Theta_1}  \circ {\bf L}\circ {\bf \bar L}^*  \circ {\bf
A_2}\circ {\bf L}={\bf \bar L}^*  \circ {\bf \Theta_1}\circ
   {\bf A_2}\circ {\bf L}={\bf \bar L}^*  \circ {\bf F}\circ {\bf L}.
$$
For $\varphi=0$  relations (11.4) are equivalence to
equalities (7.1)-(7.2) and
$${\bf F'} =  (M \textrm{cosh}\,2\theta-(e,F)\,\textrm{sinh}\, 2 \theta )+
i\{F+2e(e,F)\textrm{sinh}^2 \theta-M e \,\textrm{sinh}\, 2 \theta \}\Rightarrow$$

\emph{Relativistic formula for power  and force}:
$$
M' = \frac{{M  +v(e,F)}}{{\sqrt {1 - v^2 } }},\quad F' = (F-
e(e,F)) + e{\frac{{(e,F)-vM }}{\sqrt {1 - v^2 } }}. \eqno(11.5)$$
So, the power also depends on velocity of coordinate system. If in
initial  system it is equal to zero, but in other system it is
equal zero if only external forces are absent ( $F=0$). By this
reason  the charge conservation law is not  postulated in
traditional form
  (2.8)  for open systems, which is subjected to external influence.

\vspace{6mm}

{\large\textbf{{\textbf{12. Stress pseudotensor: Equations of
EGM-medium }}}} \vspace{4mm}

The \emph{stress pseudotensor}
 may be introduced  from formula (9.6):
$$
\sigma _{ik}^H  = - \kappa \left( {\frac{{\rho ^ H}}{{\sqrt \mu
}}\delta _{ik}  + \sqrt \mu  {\kern 1pt} {\kern 1pt} j_l^E e_{ikl}
} \right) ,\,\,\, i,k,l = 1,2,3. \eqno(12.1)$$
 It is analog of
stress tensor of liquid ($\sigma _{ik}$).

Using this pseudotensor,  (9.6) takes form, which looks
like   hydrodynamics equations:
 \[
\frac{{\partial \sigma _{ik}^H
}}{{\partial x_k }} + F_i^H  = \kappa \varepsilon \sqrt \mu \,
\frac{{\partial j_i^H }}{{\partial t}}.
\]
Here the second summand on the left-hand side is the density of
mass forces:
 \[
F_i^H  = \rho ^E E_i^{'} +  \rho ^H H_i^{'}   + j^E  \times B_i^{'}   - j^H  \times D_i^{'}
+Re (a^{'}J).
\]
However  there are not traditional index symmetries of the stress
tensor: $\sigma _{ik}  \ne \sigma _{ki} .$

Using (9.7) ,  we     may
 similarly introduce the \emph{electric stress pseudotensor}:
\[
\sigma _{ik}^E  =  - \kappa \left( {\frac{{\rho ^E
}}{{\sqrt \varepsilon  }}\delta _{ik}  - \sqrt \varepsilon  {\kern
1pt} {\kern 1pt} j_l^H e_{ikl} } \right).
\]
 By using this, the equation (9.7)  can be written as
$$
\frac{{\partial \sigma _{ik}^E }}{{\partial x_k }} + F_i^E  =
\kappa \mu \sqrt \varepsilon \, \frac{{\partial j_i^E }}{{\partial
t}}. \eqno(12.2)$$
Here the second summand on the  right-hand site
is  the density of electric forces:
 \[
F_i^E  = {\rho ^E B_i^{'}  - \rho ^H D_i^{'} } + c^{ - 1} \left( {E_i^{'}
\times j^E  + H_i^{'}  \times j^H } \right)+Im(a^{'}J).
\]
The analog of this formula is unknown for author.

\vspace{6mm}

{\large{\textbf{13 The first thermodynamics law}}}\vspace{4mm}

 We introduce the energy-impulse density for charge-current field:
$$
 0,5{\bf \Theta } \circ {\bf \Theta }^*  = \left( {
{\frac{\left\|{\rho _E }\right\|^2}{\varepsilon }}   +
{\frac{\left\|{\rho _H  }\right\|^2 }{\mu }}  + Q} \right) +
i\left( {P_J  - \sqrt {\frac{\mu }{\varepsilon }} \rho ^E j^E  -
\sqrt {\frac{\varepsilon}{\mu }} \rho ^H j^H } \right). \eqno(13.1)$$
It contains  the current energy  density:
                  \[
     Q  = 0,5\left\| J \right\|^2  = 0,5\left( {\mu \left\| {j^E }
   \right\|^2  + \varepsilon \left\| {j^H } \right\|^2 } \right),
                                 \]
where the first summand includes the Joule heat  $\left\|  {j^E }
\right\|^2  $;  the second one includes kinetic energy density of
mass current
 $\left\| {j^H } \right\|^2 $, also it contains the energy of torsional part
  of currents (magnetic current). The
vector$P_J  $ is analog of Pointing vector,  but for the current:
                                 \[
      P_J  = 0,5i\,J \times \bar J = c^{ - 1} \left[ {j^H ,j^E } \right]
                                 \]
Only if gravimagnetic and electrical currents are parallel or
one from them is equal zero, then  $P_J  = 0$.
 If  we take scalar product  in (9.5) with  $    i\bar J  $,
we get\vspace{2mm}

 \textbf{ \textit{The charge-current conservation law}}:
$$
 \kappa \left( {\partial _\tau  Q  -\,div\,P_J  +{\mathop{\rm
     Re}\nolimits} \left( {\nabla \rho ,\bar J  } \right)} \right) =
  {\mathop{\rm Im}\nolimits} \left( {F,\bar J  } \right) =c^{-1} \left( ({F^H
             ,j^H } ) + ( {F^E ,j^E }) \right).
 \eqno(13.2)$$
It is easy to see that this law is like the first thermodynamics
law. Here the sum of second and third summands in left part is
denote by $-U$. The function
                                \[
      U=\,div\,P_J  - \sqrt {\mu /\varepsilon } \left(
  {\nabla \rho ^E ,j^E } \right) -\sqrt {\varepsilon/\mu  } \left(
               {\nabla \rho ^H ,j^H } \right) \]
characterizes  the self-velocity of the change of  energy current
density of $\bf{\Theta}$-field. The right-hand site
(13.2), which depends on power of acting external forces,
can to increase or decrease this velocity.

For the free field the first thermodynamics law: $$
\partial _{\tau} Q  =U.$$

If we integrate (13.2) on $\{(S^-+S)\times(0,t)\}$ and use
Gauss formula, then the integral representation of this law may be
written as
$$\int\limits_{S^ -  } {\left( {Q (x,t) - Q (x,0)} \right)}
    dV(x) =\int\limits_0^t {dt} \int\limits_S {(P_J ,n)} \,dS(x)- $$
$$- \int\limits_0^t {dt} \int\limits_{S^ -  } {\{\varepsilon^{-1}
\left({\nabla \rho ^E ,j^E } \right)+ \mu^{-1}(\nabla \rho ^H ,j^H
)\, \} dV(x)}+ $$
$$+ c^{-1}\int\limits_0^t {dt} \int\limits_{S^ -  } \left\{(F^H,j^H ) +
 (F^E ,j^E) \right\} dV(x).$$
Here $n(x)$ is unit normal vector to boundary $S$ of the region
$S^-$ in space $R^3$. \vspace{6mm}

{\large{{\textbf{14 The total field equations  and  interaction
energy}}}}\vspace{4mm}

If there are some (N) interacting fields, generated by different
charges and currents, then Eq. (9.1) can be written as
$$
\kappa \nabla^ + {\bf \Theta }^k  + {\bf \Theta }^k  \circ
                       \sum\limits_{m \ne k} {{\bf A}^m }  = {\bf 0}
                                                           ,\,\,\,\,
             \nabla^ +  {\bf A}^k  + {\bf \Theta }^k  = {\bf 0},\quad k =
                                                       1,...,{\rm N}
\eqno(14.1)$$
$$
\nabla^ +  {\bf A}^m  \circ {\bf A}^k  + \nabla^ +  {\bf A}^k
\circ {\bf A}^m  = 0,\quad k \ne m. \eqno(14.2)$$
The total field,
as it is easy to see after summing (14.1) over $k$, is free,
because all forces are internal, also
 as in mechanics of interacting solids.

Interacting fields satisfy to the analog of the second Newton law
 (14.1), (14.2)
 and for total charge-current there is the equality:
$$
\nabla^ + {\bf \Theta } =  \nabla^ + \sum\limits_{m = 1}^M {{\bf
\Theta }^m }  ={\bf 0}. \eqno(1.  )$$

Let us  consider the laws of energy transformation  in the case of
interaction of different charges-currents. Energy-pulse for total
charge-current field reads
\[\displaylines{  {\bf \Xi }_\Theta   = 0,5{\bf \Theta } \circ {\bf \Theta }^*
= 0,5\sum\limits_{k = 1}^N {{\bf \Theta }^k }  \circ
\sum\limits_{l = 1}^N {{\bf \Theta }^{*l} }  = 0,5\left(
{\sum\limits_{k = 1}^N {{\bf \Theta }^k
 \circ {\bf \Theta }^{*k} }  + \sum\limits_{k \ne l} {{\bf \Theta }^k
  \circ {\bf \Theta }^{*l} } } \right) =  \cr    = \sum\limits_{k = 1}^N {W_\Theta  ^{(k)}
    + i\sum\limits_{k = 1}^N {P_\Theta  ^{(k)} } }  + {\bf \delta \Xi }_\Theta   \cr}\]
Here the first summand is an amount of energy-pulse of interacting charge-current.

We can introduce biquaternion of\emph{ energy-pulse interaction }.
Its real part describes energy-pulse interaction for the same name
charge and current,  but in the imagine part for different name
ones :
\[
{\bf \delta \Xi }_\Theta   = \delta W_\Theta   + i\delta P_\Theta
= \sum\limits_{k \ne l}^{} {{\bf \Xi }_\Theta  ^{kl} } ,\quad{\bf
\Xi }_\Theta  ^{kl}  = 0,5\left( {{\bf \Theta }^k  \circ {\bf
\Theta }^{*l}  + {\bf \Theta }^l  \circ {\bf \Theta }^{*k} }
\right)
\]
$${\bf \Xi }_\Theta  ^{kl}  = {\mathop{\rm Re}\nolimits} \left
( {\rho ^k \rho ^{*l}  + \left( {J^k ,J^{*l} } \right)} \right) -
i\left\{ {{\mathop{\rm Re}\nolimits} \left( {\rho ^k J^{*l}  +
 \rho ^{*l} J^k } \right) + {\mathop{\rm Im}\nolimits} \left[ {J^k ,J^{*l} } \right]} \right\},$$
or in initial notation:
$${\bf \Xi }_\Theta  ^{kl}  = \frac{{\rho
^{E(k)} \rho ^{E(l)} }}{{\sqrt {\varepsilon _k \varepsilon _l } }}
+ \frac{{\rho ^{(k)H} \rho ^{H(l)} }}{{\sqrt {\mu _k \mu _l } }} +
\sqrt {\mu _k \mu _l } \left( {j^{(k)E} ,j^{(l)E} } \right) +
\sqrt {\varepsilon _k \varepsilon _l } \left( {j^{(k)H} ,j^{(l)H}
} \right) - $$
\[
\displaylines{
   - i\left\{ {\sqrt {\frac{{\mu _l }}{{\varepsilon _k }}}
   \rho ^{(k)E} j^{(l)E}  + \sqrt {\frac{{\varepsilon _l }}
   {{\mu _k }}} \rho ^{(k)H} j^{(l)H}  + \sqrt {\frac{{\mu _k }}
   {{\varepsilon _l }}} \rho ^{(l)E} j^{(k)E}  + \sqrt {\frac{{\varepsilon _k }}
   {{\mu _l }}} \rho ^{(l)H} j^{(k)H} } \right. -  \cr
  \left. { - \sqrt {\varepsilon _k \mu _l } \left[ {j^{(l)E} \,,j^{(k)H} } \right]
  + \sqrt {\varepsilon _l \mu _k } \left[ {j^{(k)E} ,j^{(l)H} } \right]} \right\} \cr}
\]
As result  we get the conditions of energy transformation in the
case  charges-currents interaction: \emph{ energy separation } if
$\delta W_\Theta   > 0$;  \emph{energy absorption} if $\delta
W_\Theta < 0$; \emph{energy conservation } if  ${\bf \delta \Xi
}_\Theta = 0$.

\vspace{6mm}

{\large{{\textbf{ 15 Conclusion }}}}\vspace{4mm}

We considered a model of EGM-field (called \emph{A-field}),which
is founded on hypothesis on \emph{magnetic charge=mass}, that has
allowed to name such field \textit{electro-gravimagnetic}.

We used  Maxwell equations in biquaternionic form and constructed
the new biquaternionic equations for description of
charges-currents changing by their interaction. We name  these
equations as \emph{analog of Newton laws}.

 Investigation of invariance of these
equations with respect to  Lorentz transformation showed that it
is necessary to enter the \emph{scalar field of  resistance}
$a(\tau,x)$ in scalar part of biquaternion of  EGM-field tension.
One has to modify the Maxwell equations in the case, when charge
and current are subjected to influence by external field. We call
these equations  \emph{the modified Maxwell equation for open
system}.

When constructing  the equation to charge-current transformations
aside from the known gravita\-tio\-nal and electromagnetic forces
we found the presence of new forces, which is needed in
experimental motivation. Some suggestions on this cause were
presented in [3,4], where this model was offered at first, but
with charge-current conservation law  in traditional form. But
this  is true only for closed system. As it is   shown  here, for
open system we must take into account  the power of external
forces, which changes the form of  this law.

Note also that essential at building and studying  this models of
EGM-field  the algebra of biquaternions is using, without which
construction of the differential equations, describing
inter\-ac\-tion of charge and current in such form would be
practically impossible. \vspace{10mm}

{ \large{  \textbf{References}}}\\ \\
1. L.A.  Alexeyeva.  Hamilton
form of Maxwell equations and its genera\-lized solutions.
Differen\-tial equations.\textbf{39}(2003). N 6. 769-776.\\
2. L.A.  Alexeyeva. Quaternions of Hamilton form of Maxwell
equations. Mathematical journal. \textbf{3}(2003).N 4.20-24.\\
3.L.A.  Alexeyeva. About one models of electro-gravimagnetic field.
    The  interaction equa\-tions and conser\-vation laws. Mathematical journal. \textbf{4}
    (2004) N 2. 23-34. \\
4. L.A.  Alexeyeva. The  interaction equations of \emph{A}-fields
and Newton laws. Transactions of National Academy of Sci. of Rep.
Kazakhstan.
Physical and mathematical issue. 2004. N 3. 45-53. \\
5. R. Rastall.  Quaternions in relativity. Rev.
Modern Physics., \textbf{34}(1964). 820-832. \\
6. A.P.Efremov. Quaternions: algebra, geometry, and physical
  theories. Hypercomplex num\-bers in geometry and physics. \textbf{1}(2004).
   N 1. 111-127.\\
7. G.Casanova. \emph{Vector algebra.} Mir. Moscow. 1979.
(in Russian)\\
8. V.V.Kassandrov.  Biquaternion
electrodynamics and Weyl-Cartan    geometry of space-time.
    Gravitation and cosmology. \textbf{1} (1995). N 3. 216-222.\\
9. V.S.Vladimirov. \emph{Generalized functions in mathema\-tical physics. }
 Nauka. Moscow. 1976.

  \end{document}